# Study and analysis the Cu nanoparticle assisted texturization forming low reflective silicon surface for solar cell application

‡M. K. Basher[1], ‡R. Mishan[2], S. Biswas[2], ‡M. Khalid Hossain[3,*], M.A.R. Akand[1], M. A. Matin[4]

[1]Solar Cell Fabrication & Research Division, Institute of Electronics, Atomic Energy Research Establishment, Bangladesh Atomic Energy Commission, Dhaka 1349, Bangladesh.
[2] Institute of Energy, University of Dhaka, Dhaka 1000, Bangladesh.
[3]Institute of Electronics, Atomic Energy Research Establishment, Bangladesh Atomic Energy Commission, Dhaka 1349, Bangladesh
[4] Department of Glass and Ceramic Engineering, Bangladesh University of Engineering and Technology, Dhaka 1000, Bangladesh.

*E-mail: khalid.baec@gmail.com, ORCID iD: 0000-0003-4595-6367

**Abstract**

Monocrystalline silicon solar cells with photo-absorbing morphology can amplify light-trapping properties within the absorber layer and help to fabricate cost-effective solar cells. In this paper, the effect of different parameters namely temperature and time of Cu-assisted chemical etching was thoroughly investigated for the optimization of the light absorption properties. P-type monocrystalline wafers were selectively treated with $Cu(NO_3)_2.3H_2O:HF:H_2O_2$:DI water solution at $50^0C$ for five different time duration. The entire process was repeated at five different temperatures for 20min as well to study the relation between etching temperature and surface reflectance. Sonication bathing was used for the removal of the deposited Cu atoms from the surface with the variation of time and the effect was examined using energy dispersive spectroscopy (EDS). Field emission scanning electron microscopy (FESEM) and UV/VIS spectroscopy were conducted to study the surface morphology and light absorbance respectively. Inverted shapes almost similar to inverted pyramids or porous surface were found randomly on the surface of the wafer. The effect of temperature was found more significant compared to the effect of time variation. An optimum light reflectance was found at $50^0C$ for 20 min of texturization. Atomic force microscopy (AFM) of the textured sample revealed the average depth of pyramidal shape was about 1.58 µm. EDS results showed a proportional relation between time and Cu removal process, and a complete Cu atoms free textured surface after 25 min of sonication bathing. Therefore, a suitable Cu-assisted texturization technique was found, which could enable lowering the photo-reflectance below 1% without any antireflection coating.

Keywords: Inverted pyramid; metal-assisted chemical etching; texturing; silicon solar cell

## 1. Introduction

Solar energy is one of the leading sources of electricity generation because of its cost-free and infinite properties. Although there are many types of solar cells available, silicon solar cells have the lion's share because of its consistency as well as availability[1,2]. The trend of current photovoltaic cell manufacturing based on monocrystalline Si (c-Si) is increasing the overall efficiency under long-run sustainability, and reducing cost per watt so that it becomes more affordable. To elevate the conversion efficiency, different morphological transformations of solar cells have been studied [3,4]. Surface etching forms different structural shapes, for example, micron-scale textured surfaces having antireflection properties along with photoluminescence characteristics.

Among different texturization methods, the copper nanoparticles (Cu-NPs) assisted etching is getting more popular due to its simplicity, low cost, and effectiveness [5–7]. In addition, it has the flexibility to form inverted pyramid with higher surface area-to-volume ratio which facilitates the reduction of reflection losses producing higher refractive index (n) [8–11]. Thus, the higher value of n can enhance the light-trapping property of the absorber layer so that more light can be absorbed [12].

Although Cu-assisted texturization has been practised largely in recent years, the concentration of the etchant solutions [13], duration of the chemical operation and temperature range are not confirmed yet. It is observed that etching with $Cu^{2+}$/Cu-NPs solution was carried out usually for 5-30 minutes [14] and the temperature varied for $40^0C$-$90^0C$ [9,13]. This research studied the direct effect of temperature along with the time for optimizing the light-trapping property of monocrystalline silicon solar cells.

However, particles contamination due to etchant deposition has been a major concern in metal assisted chemical etching (MACE) process from the beginning as these contaminants can reduce efficiency. The strong electrostatic force makes it hard to fix this impurity. Currently, available ways of cleaning such as pressurized wet scrubs, wet chemical baths, and rotating wafer scrubbers have a tendency to damage the substrates. Ultrasonic cleaning mechanism using cavitation, vibration could be a better alternative in this situation[15,16]. Here, an efficient process of cleaning, based on ultrasonic energy from a low transducer was also applied with time variation to propose a suitable sonication bathing. This system can be associated with an adapted vacuum chunk having an acoustic wave emitter.

In this research, boron-doped monocrystalline silicon wafers were textured using $Cu^{2+}$/Cu-NPs solution to increase the roughness. To analyze the effect of temperature, the process was carried out at four different temperatures for a constant time, and to study the effect of time, the process was carried out with time variation at a constant temperature. The EDS analysis was carried out to find out the effect of the time of sonication that removes Cu-NPs completely from the surface. After that, SEM, and UV-Vis spectroscopy were used for analyzing the effect of time and temperature so that an optimum reflectance can be achieved. Finally, the surface roughness of the optimized sample was analyzed using AFM.

## 2. Experimental

Eight p-type monocrystalline silicon wafers <100> (ReneSola, China), having thickness of 200 ± 20 μm, area of 127 x 127 $mm^2$, and resistivity of 1-3 $\Omega.cm^2$ were cleaned for 10 min in 98 % pure acetone ($C_3H_6O$) at room temperature. After that for further cleaning 99% ethanol ($CH_3CH_2OH$) was used for 10 min. Then all the wafers were immersed in 5% hydrochloric acid (HCl) for 5 min to remove $SiO_2$. Next, rinsing was done using DI water to wash out the acids. Thus the Si-wafers were prepared for metal assisted chemical etching. Cu-NPs adhesion was carried out using $Cu(NO_3)_2 \cdot 3H_2O$:HF:$H_2O_2$:DI water solution. A series of samples were textured at $50^0C$ for 5min, 10min, 15min, 20min, and 25 min. The next series of the wafers were etched at $60^0C$, $70^0C$, and $80^0C$ for 20min. Sonication bath was used at 50.3 KHz and 65% $HNO_3$ for 20 min to remove the chemical used in chemical etching. A mixture of 5% (V/V) Isopropyl Alcohol and 2% KOH was used for removing $HNO_3$. Finally, all the samples were cleaned by rinsing DI water and dried by $N_2$ gas. For characterization, the FESEM (JSM-7600F, USA), Atomic force microscopy (AFM), and EDS (JSM-7600F, USA) were used. Surface reflectance was measured using UV-Vis spectrophotometer (PerkinElmer Lambda 1050, USA).

## 3. Results and discussion

### 3.1 Technique of metal deposition and formation of inverted shapes

Electrochemical reaction took place in between Si and $Cu^+$ ions through Cu-NPs-assisted anisotropic etching mechanism. An electrochemical energy difference between $Cu^+$/Cu-NPs and Si facilitates this reaction. The entire process can be described by the following reactions [3,13,17]:

Anode reaction:

Si (s)+ $2H_2O \rightarrow SiO_2 + 4H^+ + 4e^-$
$SiO_2 + 6HF \rightarrow H_2SiF_6 + 2H_2O$

Cathode reaction:

$Cu^+ + 2e^- \rightarrow Cu^0$
$H_2O_2 \rightarrow 2H_2O + 2h^+$

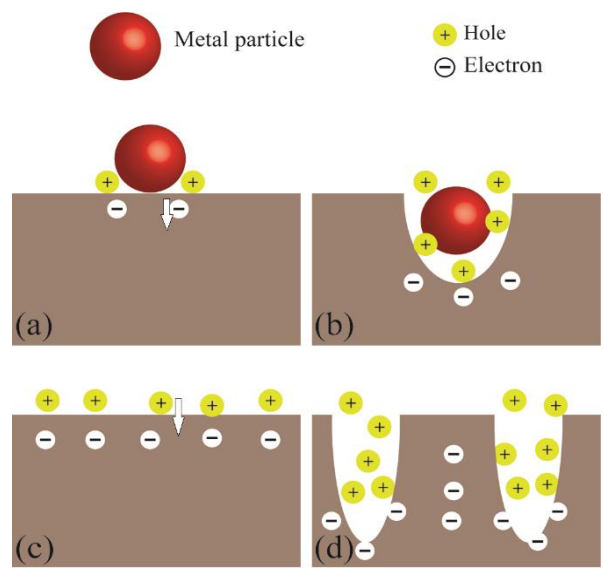

*Fig 1. Schematic representation of the mechanism of metal assisted texturization: (a) after treating with the etchant the present of Cu particle, (b) the general formation of inverted shape by producing pores, (c) the pre-stage of generating nano-pits and (d) enhanced local etching during surface evolution.*

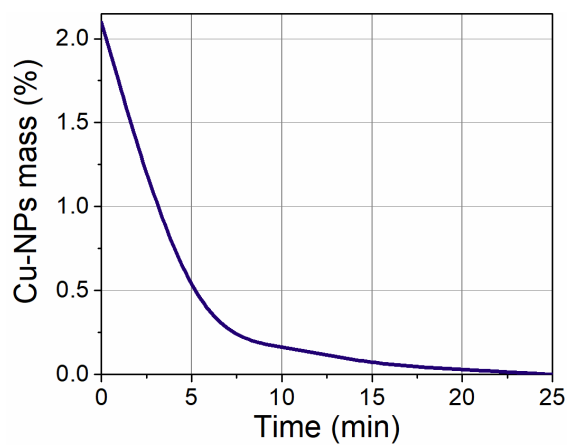

Fig 2. Reduction of Cu from the textured surface using sonication bathing treatment.

During this chemical treatment, Si atoms below the Cu-NPs are oxidized and in parallel etched by Hydrofluoric acid (HF). On the other hand, $H_2O_2$ reduction helps to deposit anisotropic Cu-NPs on the silicon (Si) substrate surface. The Cu nanoparticles get inserted into the Si substrate by grooving a pyramid like inverted shape as shown in Fig. 1(a) and (b). Thus, inverted pyramids are formed on the surface [18–20]. The holes generated from Cu into Si are stimulated repeatedly so that they could get injected to the grooved surface beneath silicon, and Cu particles deposited randomly to create a more porous Si surface with the assistance of local electric field as depicted in Fig. 1(c) and (d) [21]. The irregular surface morphology and damages of raw wafers could be a possible reason for the randomness of the distribution of inverted pyramid.

*3.2 Removal of deposited metal from the surface*

The deposited Cu-NPs were removed from the surface. Fig. 2 shows the steps of metal-etchant removal rate with respect to time. After texturization, Si surface nanopores were filled with Cu and by EDS analysis about 2.2 % copper was traced on the surface. The percentage of mass was decreased to 0.24 % after only 5 min of sonication bathing. This downward trend continued over time and decreased to 0 for 25 min of the process. The removal process worked due to the continuous formation and smash of the bubble of liquid or gas undergoing pressure changes [15].

*3.3 Effect of time variation on morphology and optical reflectance*

Visual difference between raw wafer and 5 min textured SEM images (**Fig. 3**(a)-(b)) proved distinctive morphological changes and formation of irregular inverted shapes (pyramidal and semi-pyramidal porous surface) due to Cu-assisted etching. The depth of produced inverted shapes was increased for 10 min of chemical treatment, although those were not uniform throughout the surface. For 15 min etching, more uniformly distribution of produced structures having similar depth was found. The etching ratio became high with the increase of time and thus, the depth of the inverted pyramidal or semi-pyramidal shapes was increased. Analyzing the SEM of 25 min textured sample (Fig. 3(c)), it is clear that the process of etching became slower after 20 min as the depth was decreased considerably. The UV/VIS spectroscopy analysis (Fig.4) clearly depicts that the etching time did not have any significant effect. Although texturization for only 5 to 10 min lowered the surface reflectance from approximately 31% (raw wafer) to below 1 % (0.61 % for both 5 min and 10 min textured samples) after that light absorbance did not increase that much. The proportion of light reflectance remained below 1% till 20 min after that it went high at 1.3%. However, this result is pretty low compared to a conventional upright pyramidal textured surface obtained by Ming Cao *et.al.* and Basher *et al.* [22,23].

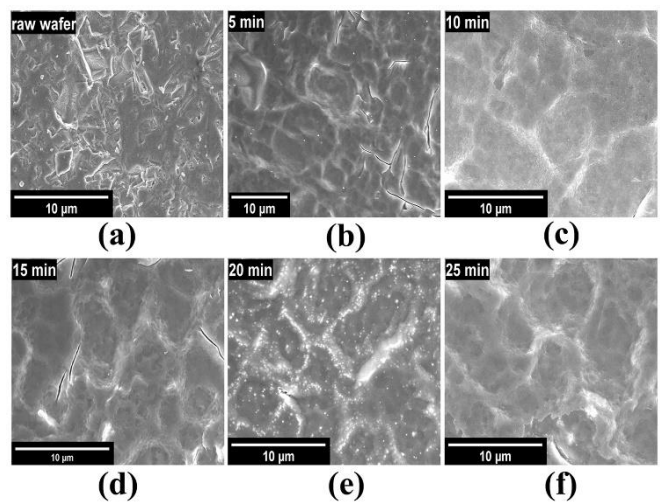

Fig 3. Surface morphology of wafers: (a) raw, (b) 5 min, (c) 10 min, (d) 15min, (e) 20min, and (f) 25min of sonication bathing at $50^0C$.

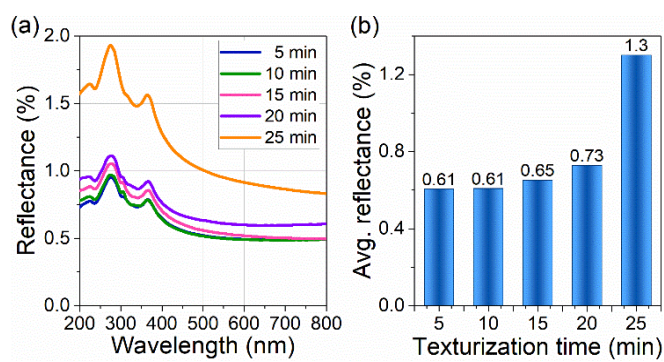

Fig 4. Optical reflectance in percentage of the inverted pyramid arrays obtained via metal-assisted etching for (a) different time and (b) average change over time.

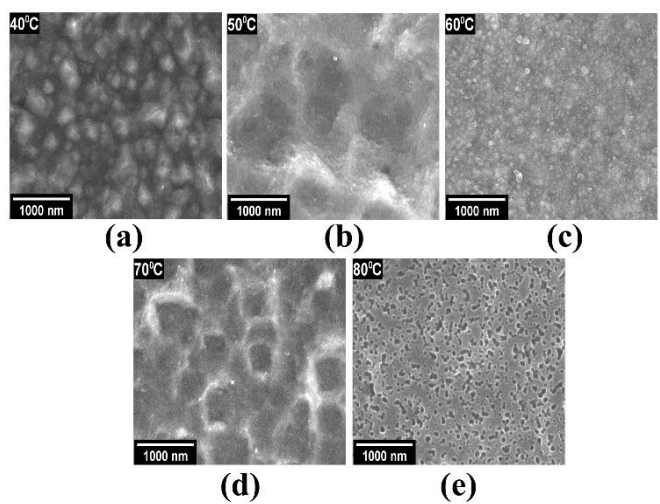

Fig 5. Surface morphology of textured wafers at a) $40^0$ C, b) $50^0$ C. c) $60^0$ C, d) $70^0$ C and e) $80^0$ C for 20 min.

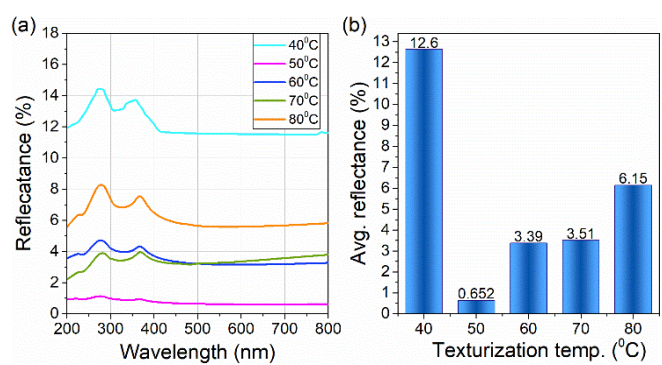

Fig 6. Reflectance in percentage of the inverted pyramid arrays obtained via metal-assisted etching for (a) different temperatures, and (b) average change of reflectance over temperatures.

### 3.4 Effect of temperature variation on morphology and optical reflectance

The temperature effect on the surface morphology (Fig.5) shows the formation of the inverted pyramids stared at $40^0$C in Fig. 5(a), and only an increase of $10^0$ C of the temperature the etching process became significantly higher, so that produced shapes were clearly visible and SEM (Fig. 5(b)). This changes decreased the ratio of reflectance from 12.64% to 0.65% because of $10^0$ C, but this did not continue for further rising of temperature. Light trapping ability decreased with the rise of temperature and surface reflectance became 3.39 %, 3.51 %, and 6.15 % for $60^0$C, $70^0$C and $80^0$C respectively (Fig 6). More double and triple bounce of light took place due to inverted pyramidal textured surface, which ensured the maximum photoabsorbance. The increasing length of light-path got absorbed on the surface and made this surface geometry as an effective and high efficiency with just 0.65% of reflectance. In addition, the unevenness morphology could also help metallization process so that better adhesion between contact metal and Si can be confirmed [13,24].

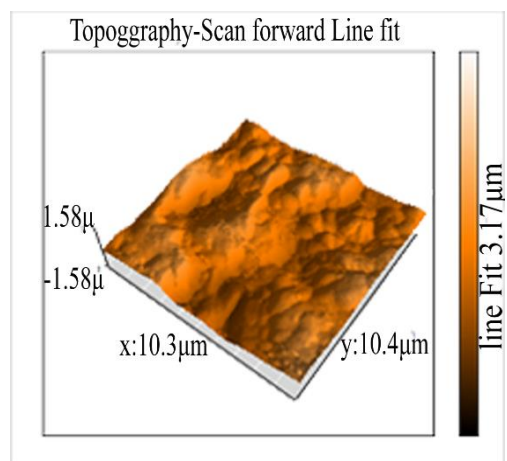

Fig 7. Surface roughness variation of 20 min textured surface at $50^0$C.

SEM images confirmed nano-pits formation into the pyramidal shapes at $80^0$C operation (Fig 5-e) even though the reflectance did not decrease for this. Incorporating time and temperature effect on the surface would be fair to suggest that at $50^0$ C for 20min produces an optimized surface reflectance of 0.652 %. In this case, the surface topology of this optimized sample obtained by AFM (**Fig.**7) showed the inverted depth of etching of about 1.58 μm although the pyramids were not well-distributed. Hence, it can be inferred that the non-uniformity of pyramidal shapes does not affect the light absorbance property of the surface significantly.

### 4. Conclusions

The impact of time and temperature variation on inverted pyramidal Cu-assisted textured surface and light-trapping property were successfully analyzed to achieve an optimized electrochemical etching technique for the fabrication of high efficiency solar cell. The variation of time did not show any major difference on the surface texture and the optical reflectance characteristics of the surface. Average optical reflectance rate was found to vary from 0.6 % to 1.3% which was significantly lower than the raw Si wafer. The reason behind this was the formation of inverted porous surface, which ensured more double and triple light-bounce, in other word better photoabsorbance. Temperature variation effect was more remarkable compared to time. At $40^0$ C temperature, surface etching was clearly observed but the inverted pyramids were not precisely formed. Only $10^0$ C increase in temperature boosted the inverted pyramid formation on the surface and at $80^0$ C nano-pits were found on the pyramidal shapes which ensured high temperature accelerated oxidation process. Average light reflectance for $40^0$, $50^0$, $60^0$, $70^0$, and $80^0$ C were 0.65%, 3.39%, 3.51%, 6.15% respectively indicate that the formation of nano-pits did not play any significant role in the increase the photoabsorbance of the surface. Possibly, these nano-pits lowered the number of double and triple light-bounces, therefore, raised the surface reflectance noticeably. Overall, the results showed that the type of shapes is less significant to achieve better light trapping properties. EDS analysis ensured the presence of Cu atoms on the textured surface as well as a gradual removal of those atoms from the surface over time in sonication bathing. For 25 min of sonication, Cu-atoms was completely removed from the wafer. Therefore, Cu-assisted etching for 15 min at $50^0$C gives the optimized reflectance 0.65%, by producing irregular inverted shapes having an average depth of 1.58μm, which may facilitate the industrial fabrication process for cost-effective and high-efficiency solar cells.

## Conflicts of interest

The authors declare that they have no competing interests.

## Acknowledgements

The authors are grateful to GCE, BUET, Bangladesh for FESEM analysis, and thankful to Institute of Energy, University of Dhaka, and IE of Bangladesh Atomic Energy Commission for supporting the entire work.

## Author contributions

MKB and MKH designed the concept. SB, RM, MKB, and MKH wrote the MS. MARK and MAM reviewed the MS.